 \documentclass[prc,showpacs,preprintnumbers,twocolumn,
                   superscriptaddress,amsmath,amssymb,floatfix]{revtex4}
\usepackage{amsfonts}
\usepackage{mathrsfs}
\usepackage{dcolumn}
\usepackage{bm}
\usepackage{longtable}
\usepackage{dsfont}
\usepackage{graphicx,epsfig,latexsym,amssymb}
\usepackage{multirow,amsmath,array,booktabs,color}
\usepackage[section]{placeins}

\newcommand{\beq}{\vspace{0.5em}\begin{equation}}
\newcommand{\eeq}{\end{equation}\vspace{0.5em}}
\newcommand{\beqn}{\vspace{0.5em}\begin{eqnarray}}
\newcommand{\eeqn}{\end{eqnarray}\par\vspace{0.5em}\noindent}

\newcommand{\beqa}{\vspace{0.5em}\begin{eqnarray*}}
\newcommand{\eeqa}{\end{eqnarray*}\par\vspace{0.5em}}
\newcommand{\bea}{\begin{array}}
\newcommand{\eea}{\end{array}}

\newcommand{\bra}[1]{\left\langle #1 \right|}
\newcommand{\ket}[1]{\left| #1 \right\rangle}

\newcommand{\bcen}{\begin{center}}
\newcommand{\ecen}{\end{center}}
\newcommand{\btab}{\begin{tabular}}
\newcommand{\etab}{\end{tabular}}
\newcommand{\bsub}{\begin{subequations}}
\newcommand{\esub}{\end{subequations}}

\newcommand{\bp}{{\mathbf{p}}}

\newcommand{\bP}{{\mathbf{P}}}

\begin{document}

\title{Configuration mixing of angular-momentum projected
triaxial relativistic mean-field wave functions}
\author{J. M. Yao}
\email{jmyao@swu.edu.cn}
\address{School of Physical Science and Technology, Southwest
          University, Chongqing 400715, China}
\address{State Key Laboratory of Nuclear Physics and Technology,
  School of Physics, Peking University, Beijing 100871, China}
\address{Physik-Department der Technischen Universit\"at M\"unchen, D-85748
 Garching, Germany}
\author{J. Meng}
\address{State Key Laboratory of Nuclear Physics and Technology,
  School of Physics, Peking University, Beijing 100871, China}
\address{School of Physics and Nuclear Energy Engineering, Beihang University, Beijing 100191, China}
\author{P. Ring}
\address{Physik-Department der Technischen Universit\"at M\"unchen, D-85748
         Garching, Germany}
\author{D. Vretenar}
\address{Physics Department, Faculty of Science, University of Zagreb, 10000 Zagreb, Croatia}

\date{\today}
\vspace{2em}
\begin{abstract}%
The framework of relativistic energy density functionals is extended
to include correlations related to the restoration of broken
symmetries and to fluctuations of collective variables.  The
generator coordinate method is used to perform configuration mixing
of angular-momentum projected wave functions, generated by
constrained self-consistent relativistic mean-field calculations for
triaxial shapes. The effects of triaxial deformation and of
$K$-mixing is illustrated in a study of spectroscopic properties of
low-spin states in $^{24}$Mg.
\end{abstract}
\maketitle


\section{Introduction}\label{Sec.I}

Among the microscopic approaches to the nuclear many-body problem,
the framework of nuclear energy density functionals (EDF) is the
only one that can presently be used over the whole nuclear chart,
from relatively light systems to superheavy nuclei, and from the
valley of $\beta$-stability to the particle drip-lines
\cite{Bender03,Vretenar05,Meng06}. Modern energy density functionals
provide the most complete and accurate description of structure
phenomena related to the evolution of shell structure in medium-mass
and heavy nuclei, e.g. the appearance of new regions of deformed
nuclei, shape coexistence and shape transitions.

In practical implementations the EDF framework is realized on two
specific levels. The simplest implementation is in terms of
self-consistent mean-field models, in which an EDF is constructed as
a functional of one-body nucleon density matrices that correspond to
a single product state -- Slater determinant of single-particle or
single-quasiparticle states. This framework can thus also be referred
to as single reference (SR) EDF. In the self-consistent mean-field
approach the many-body problem is effectively mapped onto a one-body
problem, and the exact EDF is approximated by a functional of powers
and gradients of ground-state nucleon densities and currents,
representing distributions of matter, spins, momentum and kinetic
energy. In principle the SR nuclear EDF can incorporate short-range
correlations related to the repulsive core of the inter-nucleon
interaction, and long-range correlations mediated by nuclear
resonance modes. The static SR EDF is characterized by symmetry
breaking -- translational, rotational, particle number, and can only
provide an approximate description of bulk ground-state properties.
To calculate excitation spectra and electromagnetic transition rates
in individual nuclei, it is necessary to extend the SR EDF framework
to include collective correlations related to the restoration of
broken symmetries and to fluctuations of collective coordinates.
Collective correlations are sensitive to shell effects, display
pronounced variations with particle number, and cannot be
incorporated in a SR EDF. On the second level that takes into account
collective correlations through the restoration of broken symmetries
and configuration mixing of symmetry-breaking product states, the
many-body energy takes the form of a functional of all transition
density matrices that can be constructed from the chosen set of
product states. This level of implementation is also referred to as
multireference (MR) EDF framework.

In recent years several accurate and efficient models and algorithms
have been developed that perform the restoration of symmetries broken
by the static nuclear mean field, and take into account fluctuations
around the mean-field minimum. The most effective approach to
configuration mixing calculations is the generator coordinate method
(GCM) \cite{Ring80,Blaizot86}. With the simplifying assumption of
axial symmetry, GCM configuration mixing of angular-momentum, and
even particle-number projected quadrupole-deformed mean-field states,
has become a standard tool in nuclear structure studies with Skyrme
energy density functionals \cite{Valor00,Bender03}, the
density-dependent Gogny force \cite{Guzman02npa}, and relativistic
density functionals \cite{Niksic06I,Niksic06II}. A variety of
structure phenomena have been analyzed using this approach. For
instance,  the structure of low-spin deformed and superdeformed
collective states~\cite{Guzman00,Bender03prc,Bender04prc}, shape
coexistence in Kr and Pb isotopes~\cite{Guzman04prc,Bender06prc},
shell closures in the neutron-rich Ca, Ti and Cr
isotopes~\cite{Rodriguez07} and shape transition in Nd
isotopes~\cite{Niksic07PRL,Rodriguez08}.

Much more involved and technically difficult is the description of
intrinsic quadrupole modes including triaxial deformations. Intrinsic
triaxial shapes are essential for the interpretation of interesting
collective modes, such as chiral rotations
\cite{Frauendorf97,Grodner06} and wobbling motion~\cite{Odegard01}.
The inclusion of triaxial shapes can dramatically reduce barriers
separating prolate and oblate minima, leading to structures that are
soft or unstable to triaxial distortions~\cite{Cwiok05}. Such a
softness towards dynamical $\gamma$-distortions will give rise to the
breakdown of the $K$-selection rule in electromagnetic transitions of
high-spin isomers~\cite{Chowdhury88}. It may also has important
influence on the electric monopole transition strength $B(E0:
0^+_2\rightarrow0^+_1)$~\cite{Wiedeking08}.

Only very recently a fully microscopic three-dimensional GCM model
has been introduced~\cite{Bender08}, based on Skyrme mean-field
states generated by triaxial quadrupole constraints that are
projected on particle number and angular momentum and mixed by the
generator coordinate method. This method is actually equivalent to a
seven-dimensional GCM calculation, mixing all five degrees of freedom
of the quadrupole operator and the gauge angles for protons and
neutrons. In this work we develop a model for configuration mixing of
angular-momentum projected triaxial relativistic mean-field wave
functions. In the first part, reported in Ref.~\cite{Yao09amp}, we
have already considered three-dimensional angular-momentum projection
(3DAMP) of relativistic mean-field wave functions, generated by
constrained self-consistent mean-field calculations for triaxial
quadrupole shapes. These calculations were based on the relativistic
density functional PC-F1~\cite{Burvenich02},  and pairing
correlations were taken into account using the standard BCS method
with both monopole and zero-range $\delta$ interactions. Correlations
related to the restoration of rotational symmetry broken by the
static nuclear mean field, were analyzed for several Mg isotopes.
Here we extend the model of Ref.~\cite{Yao09amp}, and perform GCM
configuration mixing of 3DAMP relativistic mean-field wave functions.

In Section~\ref{Sec.II} we introduce the model, briefly outline the
relativistic point-coupling model which will be used to generate
mean-field wave functions, and describe in detail the procedure of
configuration mixing of angular momentum projected wave functions. In
Section~\ref{Sec.III} the 3DAMP+GCM model is tested in illustrative
calculations of the low-energy excitation spectrum of $^{24}$Mg.
Section \ref{Sec.IV} summarizes the results of the present
investigation and ends with an outlook for future studies.

\section{The 3DAMP+GCM model}
 \label{Sec.II}
The generator coordinate method (GCM) is based on the assumption
that, starting from a set of mean-field states $\ket{\Phi (q)}$
which depend on a collective coordinate $q$, one can build
approximate eigenstates of the nuclear Hamiltonian
\begin{equation}
\ket{\Psi_\alpha} =  \int d q {f_\alpha(q)\ket{\Phi (q)}}\;.
\label{GCM-state}
\end{equation}
Detailed reviews of the GCM can be found in \cite{Ring80,Blaizot86}. In the present study
the basis states $\ket{\Phi (q)}$ are Slater determinants of single-nucleon states generated
by self-consistent solutions of constrained relativistic mean-field
(RMF) + BCS equations. To be able to compare theoretical predictions with data, it is
of course necessary to construct states
with good angular momentum. Thus the trial angular-momentum projected GCM collective
wave function $\vert\Psi^{JM}_\alpha\rangle$, an eigenfunction of $\hat J^2$ and $\hat J_z$, with
eigenvalues $J(J+1)\hbar^2$ and $M\hbar$, respectively, reads
 \beqn
 \label{TrialWF}
 \vert\Psi^{JM}_\alpha\rangle
  &=&  \int d q \sum_{K\geq0}f^{JK}_{\alpha}(q)\frac{1}{(1+\delta_{K0})}\vert JMK+,q\rangle
 \eeqn
where $\alpha=1,2,\cdots$ labels collective eigenstates for a given
angular momentum $J$. The details of the 3D angular-momentum
projection are given in Ref.~\cite{Yao09amp}, here we only outline
the basic features. Because of the $D_2$ and time-reversal symmetry
of a triaxially deformed even-even nucleus, the projection of the
angular momentum $J$ along the intrinsic $z$-axis ($K$  in
Eq.~(\ref{TrialWF}) ) takes only non-negative even values: \beqn
 K=\left\{\begin{array}{ccc}
 0,2,\cdots, J   & {\rm for} & J\; {\rm mod}\; 2 =0 \\
 2,4,\cdots, J-1 & {\rm for} & J\; {\rm mod}\; 2 =1 \\
 \end{array}
 \right.
 \eeqn
 The basis states $\vert JMK+,q\rangle$ are projected
 from the intrinsic wave functions $\vert\Phi(q)\rangle$:
 \beq
  \vert JMK+,q\rangle
    =[\hat P^J_{MK}+(-1)^J\hat P^J_{M-K}]\vert\Phi(q)\rangle,
 \eeq
 where $\hat P^J_{MK}$ is the angular-momentum projection operator:
  \beqn
  \hat P^J_{MK}
  =\frac{2J+1}{8\pi^2}\int d\Omega D^{J\ast}_{MK}(\Omega) \hat R(\Omega) \;.
  \eeqn
 $\Omega$ denotes the set of three Euler angles: ($\phi, \theta, \psi$), and
 $d\Omega=d\phi \sin\theta d\theta d\psi$. $D^J_{MK}(\Omega)$ is the Wigner $D$-function,
 with the rotational operator chosen in the notation of Edmonds~\cite{Edmonds57}:
 $ \hat R(\Omega)=e^{i\phi\hat J_z}e^{i\theta\hat J_y}e^{i\psi\hat J_z}$.
 The set of intrinsic wave functions $\vert\Phi(q)\rangle$, with the generic notation
 for quadrupole deformation parameters $q\equiv(\beta,\gamma)$, is generated
 by imposing constraints on the axial $q_{20}$ and triaxial $q_{22}$ mass quadrupole
 moments in self-consistent RMF+BCS calculations. These moments are related
 to the Hill-Wheeler~\cite{Hill53} coordinates $\beta$ ($\beta>0$) and $\gamma$ by
 the following relations:
 \bsub%
 \beqn%
 q_{20}&=& \sqrt{\frac{5}{16\pi}}\langle 2z^2-x^2-y^2\rangle = \frac{3}{4\pi}AR^2_0\beta\cos
 \gamma,~~~~~~~~~~~\\
 q_{22}&=& \sqrt{\frac{15}{32\pi}}\langle x^2-y^2\rangle = \frac{3}{4\pi}AR^2_0 \frac{1}{\sqrt{2}}\beta\sin\gamma,
 \eeqn
 \esub%
 where $R_0=1.2A^{1/3}$ fm. The total mass quadrupole moment $q_m$ reads:
 \beq
 \label{Q-moment}%
 q_m= \sqrt{\frac{16\pi}{5}} \sqrt{q^2_{20}+2q^2_{22}}.
 \eeq

The calculation of single-nucleon wave functions, energies and occupation factors starts
with the choice of the energy density functional (EDF). As in our
previous analysis on collective correlations in axially deformed nuclei
\cite{Niksic06I,Niksic06II}, and in the first part of this work \cite{Yao09amp}, the
present illustrative calculation is based on
the relativistic functional PC-F1 (point-coupling Lagrangian) \cite{Burvenich02}:
 \begin{eqnarray}
 {{E}}_{\rm RMF}
 &=& \int d{\bm r }~{\mathcal{E}_{\rm RMF}}(\bm{r}) \\
 &=& \sum_k{\int d\bm{r}~v_k^2
  ~{\bar{\psi}_k (\bm{r}) \left( -i\bm{\gamma}\bm{\nabla} + m\right )\psi_k(\bm{r})}}\nonumber \\
 &+& \int d{\bm r }~{\left(\frac{\alpha_S}{2}\rho_S^2+\frac{\beta_S}{3}\rho_S^3 +
    \frac{\gamma_S}{4}\rho_S^4+\frac{\delta_S}{2}\rho_S\triangle \rho_S \right.}\nonumber \\
 &+&  {\left.\frac{\alpha_V}{2}j_\mu j^\mu + \frac{\gamma_V}{4}(j_\mu j^\mu)^2 +
     \frac{\delta_V}{2}j_\mu\triangle j^\mu \right.} \nonumber \\
 &+& \left. \frac{\alpha_{TV}}{2}j^{\mu}_{TV}(j_{TV})_\mu+\frac{\delta_{TV}}{2}
    j^\mu_{TV}\triangle  (j_{TV})_{\mu}\right.\nonumber \\
    &+&\frac{\alpha_{TS}}{2}\rho_{TS}^2
   \left.+\frac{\delta_{TS}}{2}\rho_{TS}\triangle
      \rho_{TS} +e\frac{1-\tau_3}{2}\rho_V A^0
 \right) ,
\nonumber
\label{EDF}
\end{eqnarray}
 where $\psi_k(\bm{r})$ denotes a Dirac spinor.
 The local isoscalar and isovector densities and currents
 \bsub\begin{eqnarray}
 \label{dens_1}
 \rho_{S}({\bm r}) &=&\sum_{k>0} v_k^2 ~\bar{\psi}_{k}({\bm r}) \psi _{k}({\bm r})~,  \\
 \label{dens_2}
 \rho_{TS}({\bm r}) &=&\sum_{k>0} v_k^2 ~ \bar{\psi}_{k}({\bm r})\tau_3\psi _{k}^{{}}({\bm r})~,  \\
 \label{dens_3}
 j^{\mu}({\bm r}) &=&\sum_{k>0} v_k^2 ~\bar{\psi}_{k}({\bm r}) \gamma^\mu\psi _{k}^{{}}({\bm r})~,  \\
 \label{dens_4}
 j^{\mu}_{TV}({\bm r}) &=&\sum_{k>0} v_k^2 ~\bar{\psi}_{k}({\bm r}) \gamma^\mu
 \tau_3 \psi _{k}^{{}}({\bm r})\;,
 \end{eqnarray}\esub
 are calculated in the {\it no-sea} approximation, i.e., the
 summation runs over all occupied states in the Fermi sea.
The occupation factors $v_k^2$ of each orbit are determined in the
simple BCS approximation, using a $\delta$-pairing force. The pairing
contribution to the total energy is given by
 \beq
 \label{PairingDFT}
 E_{\rm pair}[\kappa,\kappa^*]
 = - \sum_{\tau=n,p} \dfrac{V_\tau}{4}\int d^3r\kappa^\ast_\tau(\bm{r}) \kappa_\tau(\bm{r}).
 \eeq
 where $V_\tau$ is a constant pairing strength, and the
 pairing tensor $\kappa(\bm{r})$ reads
 \beq
  \kappa(\bm{r})
  =-2\sum_{k>0}f_ku_kv_k\vert\psi_k(\bm{r})\vert^2.
 \eeq
 The pairing window is constrained with smooth cutoff factors $f_k$, determined
 by a Fermi function in the single-particle energies $\epsilon_k$:
 \beq%
 \label{Weight}
 f_k  =\frac{1}{1+\exp[( \epsilon_k- \epsilon_F-\Delta E_\tau)/\mu_\tau]} \;.
\eeq%
 $\epsilon_F$ is the chemical potential determined by the constraint on
 average particle number: $\langle \Phi(q)\vert \hat N_\tau \vert \Phi(q) \rangle = N_\tau$.
 The cut-off parameters $\Delta E_\tau$ and $\mu_\tau=\Delta E_\tau/10$ are chosen
 in such a way that $2 \sum_k f_k = N_\tau +1.65N^{2/3}_\tau$, where
 $N_\tau$ is the number of neutrons (protons)~\cite{Bender00p}.

The weight functions $f^{JK}_{\alpha}(q)$ in the collective wave function Eq.~(\ref{TrialWF})
are determined from the variation:
\begin{equation}
 \delta E^{J} =
 \delta \frac{\bra{\Psi_\alpha^{JM}} \hat{H} \ket{\Psi_\alpha^{JM}}}
            {\bra{\Psi_\alpha^{JM}}\Psi_\alpha^{JM}\rangle} = 0 \; ,
\label{variational}
\end{equation}
i.e., by requiring that the expectation value of the energy is
stationary with respect to an arbitrary variation $\delta
f_{\alpha}^{JK}$. This leads to the  Hill-Wheeler-Griffin (HWG)
integral equation:
 \beq
 \label{HWEq}
 \int dq^\prime\sum_{K^\prime\geq0}
 \left[\mathscr{H}^J_{KK^\prime}(q,q^\prime)
 - E^J_\alpha\mathscr{N}^J_{KK^\prime}(q,q^\prime)\right]
  f^{JK^\prime}_\alpha(q^\prime)=0,
 \eeq
 where $\mathscr{H}$ and $\mathscr{N}$ are the angular-momentum projected GCM
 kernel matrices of the Hamiltonian and the norm, respectively.
 With the generic notation  $\mathscr{O} \equiv \mathscr{N}$ or
 $\mathscr{H}$, the expression for the kernel reads:
 \beqn
 \label{OverlapK}
 \mathscr{O}^J_{KK^\prime}(q,q^\prime)
 &=&\Delta_{KK^\prime}
     [O^J_{KK^\prime}(q,q^\prime)
      +(-1)^{2J}O^J_{-K-K^\prime}(q,q^\prime)\nonumber\\
  &+&  (-1)^JO^J_{K-K^\prime}(q,q^\prime)
      +(-1)^JO^J_{-KK^\prime}(q,q^\prime)],\nonumber\\
 \eeqn
 where for the operator $\hat O \equiv 1$ or $\hat H$:
{\beqn
 \label{Integration1}
 O^J_{KK^\prime}(q,q^\prime)
 = \langle\Phi(q)\vert \hat O\hat{P}^J_{KK^\prime}\vert\Phi(q^\prime)\rangle ,
\eeqn
 and $\Delta_{KK^\prime}=1/[(1+\delta_{K0})(1+\delta_{K^\prime0})]$.

The overlap $\langle\Phi(q)\vert \hat H\hat
R\vert\Phi(q^\prime)\rangle$ can be evaluated in coordinate space,
and we rewrite the hamiltonian kernel $H^J_{KK^\prime}(q,q^\prime)$
in the following form:
 \beq
 H^J_{KK^\prime}(q,q^\prime)
 =\int d\bm{r}  H^J_{KK^\prime}(\bm{r};q,q^\prime) \;,
 \eeq
 where
 \beq
 H^J_{KK^\prime}(\bm{r};q,q^\prime)
 = \frac{2J+1}{8\pi^2} \int d\Omega D^{J\ast}_{KK^\prime}
  {\cal H}(\bm{r};q,q^\prime;\Omega)n(q,q^\prime;\Omega)
 \eeq
 The norm overlap $n(q,q^\prime;\Omega)$ is
 defined by:
 \beq
 n(q,q^\prime;\Omega)
 \equiv\langle\Phi(q)\vert \hat R(\Omega)\vert\Phi(q^\prime)\rangle \; .
 \eeq
The calculation of the overlap matrix elements ${\cal
H}(\bm{r};q,q^\prime;\Omega)$ requires the explicit form of $\hat
H$. So far we have implicitly assumed that the system is described
by a Hamiltonian. However, for energy density functionals this is
strictly valid only if the density dependence can be expressed as a
polynomial of $\rho$. By using product wave functions, a density
functional can formally be derived from a Hamiltonian that contains
many-body interactions. A prescription based on the generalized Wick
theorem~\cite{Balian69} states that the Hamilton overlap matrix
elements have the same form as the mean field functional,  with the
intrinsic single particle density matrix elements replaced by the
corresponding transition density matrix elements~\cite{Onishi66}. In
this work we employ the relativistic point-coupling model
PC-F1~\cite{Burvenich02}, which contains powers of the scalar
density $\rho_S$ up to fourth order, and therefore the above
prescription can be applied. For a detailed discussion of open
problems we refer the reader to Ref.~\cite{Duguet2010}, and
references cited therein.

Consequently, ${\cal H}(\bm{r};q,q^\prime;\Omega)$ has the same form as the
mean-field functional ${\mathcal{E}_{RMF}}(\bm{r})$ in
Eq.~(\ref{EDF}) provided the {\em intrinsic} densities and currents
are replaced by {\em transition} densities and currents. Further
details about the calculation of the norm overlap
$n(q,q^\prime;\Omega)$ and transition EDF ${\cal
H}(\bm{r};q,q^\prime;\Omega)$ can be found in Ref.~\cite{Yao09amp}.

The basis states $\ket{\Phi(q)}$ are not eigenstates of the proton
and neutron number operators $\hat{Z}$ and $\hat{N}$. The adjustment
of the Fermi energies in a BCS calculation ensures only that the
average value of the nucleon number operators corresponds to the
actual number of nucleons. It follows that the wave functions
$\ket{\Psi_\alpha^{JM}}$ are generally not eigenstates of the nucleon
number operators and, moreover, the average values of the nucleon
number operators are not necessarily equal to the number of nucleons
in a given nucleus. This happens because the binding energy increases
with the average number of nucleons and, therefore, an unconstrained
variation of the weight functions in a GCM calculation will generate
a ground state with the average number of protons and neutrons larger
than the actual values in a given nucleus. In order to restore the
correct mean values of the nucleon numbers, we follow the standard
prescription \cite{Hara82,Bonche90}, and modify the HWG equation by
replacing ${\cal H}(\bm{r};q,q^\prime;\Omega)$ with
 \beqn
 {\cal H}^\prime(\bm{r};q,q^\prime;\Omega)
 &=&{\cal H}(\bm{r};q,q^\prime;\Omega)
 - \lambda_p[Z(\bm{r};q,q^\prime;\Omega)-Z_0]\nonumber\\
    &&
 - \lambda_n[N(\bm{r};q,q^\prime;\Omega)-N_0],
 \eeqn
 where $Z_0$ and $N_0$ are the desired proton and neutron numbers, respectively.
 $Z(\bm{r};q,q^\prime;\Omega)$ and $N(\bm{r};q,q^\prime;\Omega)$ are the transition
 vector densities in $\bm{r}$-space for protons and neutrons, respectively.
 The Lagrange parameters $\lambda_{\tau=p,n}$
 are in principle determined in such a way that each AMP GCM collective state
 has the correct average particle number.
 In that case, however, the Lagrange parameters $\lambda_\tau$ will be state dependent and,
 as a consequence, the orthonormality of the states $\vert\Psi^{JM}_\alpha\rangle$s is no longer
 guaranteed. In Ref.~\cite{Bonche90} a simple ansatz was introduced for a
 state-independent value of the  Lagrange parameter, that is the value of $\lambda_{\tau=p,n}$
 was chosen to be the mean BCS Fermi energy, determined by averaging over the collective
 variable $q$. The average particle numbers  in the resulting AMP GCM states
 differ only slightly from the desired correct values. In the present model we
 take the same $\lambda_\tau$ values as those in the mean-field calculation,
 i.e. $\lambda_\tau(q)$ for the diagonal terms ($q^\prime=q$),
 and $[\lambda_{\tau}(q)+\lambda_{\tau}(q^\prime)]/2$ for the off-diagonal
 ones ($q^\prime\neq q$) in ${\cal H}^\prime (\bm{r};q,q^\prime;\Omega)$.
 We find that with this prescription the average particle numbers for low-lying excitation
 states are in excellent agreement with those obtained by taking the
$\lambda_\tau$ value averaged over the collective variable $q$.
\begin{figure*}[]
  \centering
  \includegraphics[width=16cm]{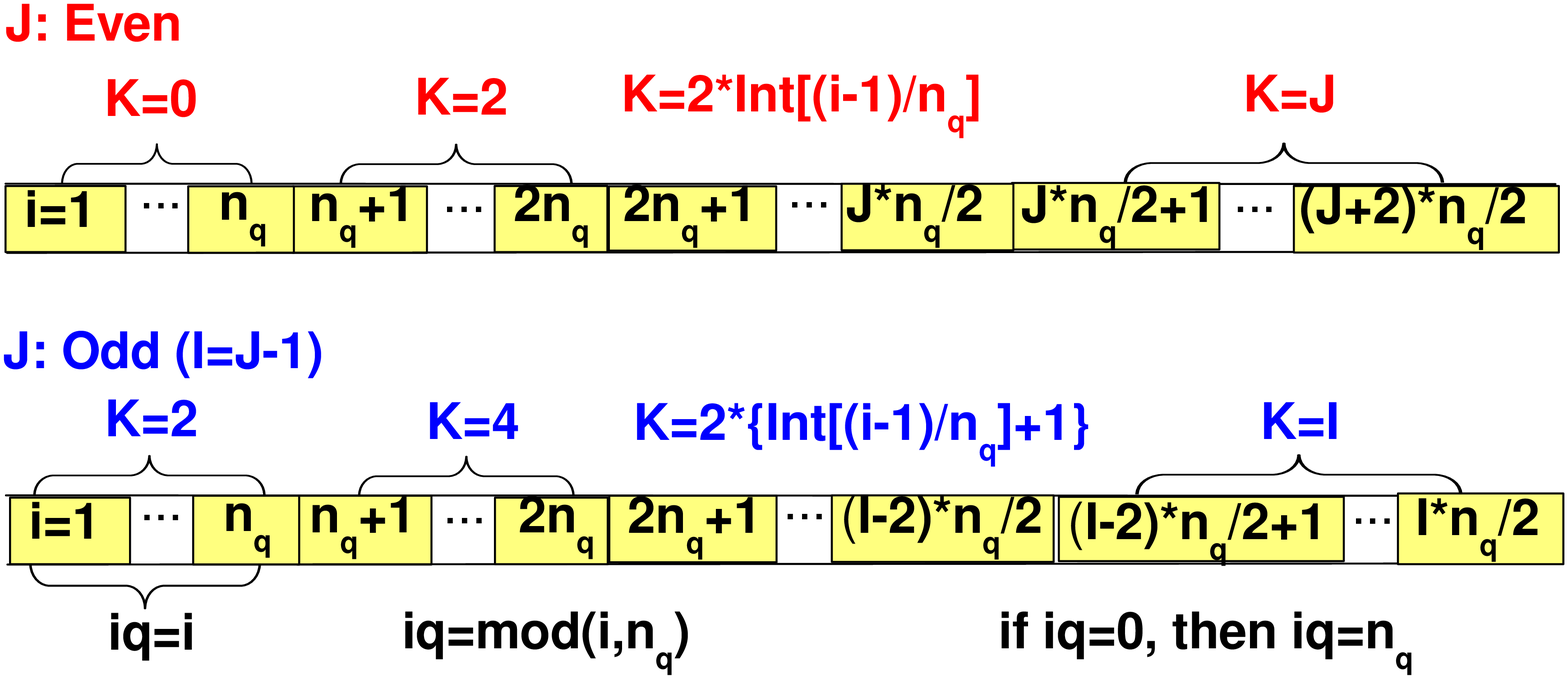}
   \caption{ (Color online) Distributions of the $K$ quantum number (K) and $q$ values (iq) in
    the full $K\bigotimes q$ space. The dimension of the direct product of the $K$-subspace and
 the $q$-subspace is $D=(J+2)n_q/2$ for even $J$, or $D=(J-1)n_q/2$ for odd $J$. $n_q$ is the number of points on the
 mesh in $q$-space, and $J$ is the total angular momentum.}
  \label{fig1}
 \end{figure*}

The domain of quadrupole deformation parameters $q\equiv(\beta,\gamma)$ is discretized,
and  the HWG integral equation is transformed into a matrix eigenvalue equation.
 The corresponding kernels $\mathscr{O}^J_{KK^\prime}(q,q^\prime)$ have to
 be calculated between all pairs of mesh points in $q$ space. In the current version
 of the model the full space  $K\bigotimes q$ is a direct product of the $K$-subspace and
 the $q$-subspace, with dimension $D=(J+2)n_q/2$ for
 even $J$ or $D=(J-1)n_q/2$ for odd $J$. $n_q$ is the number of points on the
 mesh in $q$-space, and $J$ the total angular momentum. Correspondingly,
 the kernels $\mathscr{O}^J_{KK^\prime}(q,q^\prime)$ $\longrightarrow$
 $\mathscr{O}^J (i,j)$. The quantum number $K$ and the value of $(\beta,\gamma)$
 at each point of the full space $K\bigotimes q$ can be determined
 as shown in Fig.~\ref{fig1}.

 The first step in the solution of the HWG matrix eigenvalue equation is the diagonalization of
 the norm overlap kernel $\mathscr{N}^J(i,j)$
 \beq
  \sum_j \mathscr{N}^J(i,j)u^J_k(j)=n^J_ku^J_k(i).
 \eeq
Since the basis functions $\vert\Phi(q)\rangle$ are not linearly
independent, many of the eigenvalues $n^J_k$ are very close to zero.
They correspond to ``high momentum" collective components, i.e., the
corresponding eigenfunctions $u^J_k(i)$
 are rapidly oscillating in the $q$-space but carry very little physical information.
However, due to numerical uncertainties, their contribution to %
the matrix elements of the collective Hamiltonian~(\ref{Hcoll})  can
be large, and these states should be removed from the basis.
Therefore, a small positive constant $\zeta$ is introduced so that
states with $n^J_k/n^J_{\textrm max} < \zeta$ are excluded from the
GCM basis, where $n^J_{\textrm max}$ is the largest eigenvalue of
the norm kernel. From the remaining states, also called ``{\em
natural states}", one builds the collective
Hamiltonian
 \begin{equation}
 \label{Hcoll}
 \mathcal{H}^{J}_{kl} = \frac{1}{\sqrt{n^J_k}}\frac{1}{\sqrt{n^J_l}}
  \sum_{i,j}{u^J_k(i) \mathscr{H}^J (i,j)u^J_l(j)}\;,
 \end{equation}
 which is subsequently diagonalized
 \begin{equation}
 \label{CollectiveEQ}
 \sum_{l}\mathcal{H}^{J}_{kl}g_l^{J\alpha}
 = E^J_{\alpha}g_k^{J\alpha} \;.
 \end{equation}
 The solution of Eq.~(\ref{CollectiveEQ}) determines both the
 energies $E^J_\alpha$ and the amplitudes $f^{JK}_{\alpha}(q)$ of collective states
 with good angular momentum
 $\vert \Psi^{JM}_{\alpha}\rangle$
 \beq
  \displaystyle
  f^{JK}_{\alpha}(q)
   = \sum\limits_{k}\frac{g_k^{J\alpha}}{\sqrt{n^J_k}}u^J_k(i).
 \eeq
 The weight functions $f^{JK}_{\alpha}(q)$ are not orthogonal and
 cannot be interpreted as collective wave functions for the deformation
 variables. The collective wave functions $g^J_\alpha(i)$ are calculated from
 the norm overlap eigenstates:
 \beq
 \label{probability}
 g^J_\alpha(i) =\sum\limits_{k}g_k^{J\alpha}u^J_k(i),
 \eeq
  $g^J_\alpha(i)$ are orthonormal
 and, therefore, $\vert g^J_\alpha(i)\vert^2$ can be
 interpreted as a probability amplitude.

The center-of-mass (c.m.) correction is defined by:
 \beqn
&&\langle E_{\mathrm{cm}}\rangle(J_{\alpha})
 =\langle\Psi^{JM}_{\alpha}\vert\dfrac{\hat{\mathbf{P}}_{\mathrm{cm}}^{2}}{2mA}
 \vert\Psi^{JM}_{\alpha}\rangle\nonumber\\
 &=&\sum_{i
 j}\sum_{KK^\prime}f^{JK\ast}_{\alpha}(i)f^{JK^\prime}_{\alpha}(j)
 \frac{1}{2mA}\langle\Phi(q_i)|\hat{\mathbf{P}}_{\mathrm{cm}}^{2}P_{KK^{\prime}}^{J}|\Phi(q_j)\rangle \;.\nonumber\\
 \eeqn
The projected overlap matrix elements
$\langle\Phi(q_i)|\hat{\mathbf{P}}_{\mathrm{cm}}^{2}P_{KK^{\prime}}
^{J}|\Phi(q_j)\rangle$ are treated in zeroth order of the Kamlah
approximation, i.e. considering the fact that
$\langle\Phi(q_i)|\Phi(q_j)\rangle$ is sharply peaked at $q_i=q_j$,
the projected matrix elements are approximated by the unprojected
ones~\cite{Ring80}, and
\begin{eqnarray}
 \frac{1}{2mA}\langle\Phi(q_i)|\hat{\mathbf{P}}_{\mathrm{cm}}^{2}
 \hat P_{KK^{\prime}}^{J}|\Phi(q_j)\rangle
 \approx \mathscr{N}^J_{KK^{\prime}}(q_i,q_j) E_{\mathrm{cm}}(q_i)\;,\nonumber\\
\end{eqnarray}
were $E_{\mathrm{cm}}(q)$ is the c.m. correction evaluated for the
intrinsic wave functions $\vert\Phi(q)\rangle$,
\begin{equation}
  \label{Eq:Ecm}
 E_{\mathrm{cm}}(q)
 =\frac{1}{2mA}\langle\Phi(q)|\hat{\mathbf{P}}_{\mathrm{cm}}^{2}|\Phi(q)\rangle \;,
\end{equation}
 where $m$ is the nucleon mass, and $A$ is the number of nucleons.
 $\hat \bP_{\rm cm}=\sum_i^A \hat \bp_i$ is the total momentum.
 The energy of the collective state  $\vert \Psi^{J}_{\alpha}\rangle$ is,
 therefore, given by
 \beq
  \mathbb{E}(J^+_\alpha)
  = E^J_\alpha  + \langle E_{\rm cm}\rangle(J_\alpha) \; .
 \eeq

 Once the amplitudes $f^{JK}_{\alpha}(q)$ of nuclear collective wave functions
 $\vert \Psi^{JM}_{\alpha}\rangle$ are known, it is straightforward to calculate
 all physical observables, such as the electromagnetic transition probability,
 spectroscopic quadrupole moments and the average particle number.
 The $B(E2)$ probability for a transition from an initial state
 $(J_i,\alpha_i)$ to a final state $(J_f,\alpha_f)$ is defined by
 \beqn
 && B(E2;J_i,\alpha_i\rightarrow J_f,\alpha_f)\nonumber\\
 &=& \frac{e^2}{2J_i+1}
        \left|\sum_{q_f,q_i}\langle J_f,q_f\vert\vert \hat Q_{2}\vert\vert J_i,q_i \rangle
        \right|^2 \; ,
 \eeqn
Using the generalized Wigner-Eckart theorem for the
spherical tensor operator $\hat Q_{\lambda\mu}$%
  \beqn
  \hat P^{J}_{KM}\hat{Q}^{}_{\lambda\mu} \hat P^{J^\prime}_{M^{\prime}K^\prime}
  =C^{J\,M}_{J^{\prime}M^\prime\lambda\mu}\sum_{\bar{K}\mu'}
    C^{J\,K}_{J^{\prime}\bar{K}\lambda\mu^\prime}
   \hat{Q}^{}_{\lambda\mu^\prime} \hat P^{J^\prime}_{\bar{K}K^\prime}
  \eeqn
and the relation
\beq%
\hat{P}^J_{MK}\hat{P}^{J^\prime}_{M^{\prime}K^\prime}=%
\delta_{JJ^\prime}\delta_{KM^\prime}\hat{P}^J_{MK^\prime}%
\eeq%
for projection operators~\cite{Ring80}, one obtains for the reduced
matrix element $\langle J_f,q_f\vert\vert \hat Q_{2}\vert\vert
J_i,q_i\rangle$:%
\beqn
 \label{Integration2}
 \langle J_f,q_f\vert\vert \hat Q_{2}\vert\vert J_i,q_i\rangle
 ={\hat J}_f
 \sum_{K_iK_f}f^{\ast J_fK_f}_{\alpha_f}(q_f)f^{J_iK_i}_{\alpha_i}(q_i)~~~~~\\
 \times
 \sum_{\mu K^\prime}(-1)^{J_f-K_f}
 \left(\begin{array}{ccc}
   J_f  &  2         &J_i \\
   -K_f & \mu        &K^\prime \\
 \end{array}
 \right) Q_{2\mu}(K^\prime,K_i;q_f,q_i)
\nonumber
\eeqn%
with ${\hat J}_f=2J_f+1$, $f^{JK}_{\alpha}(q) =
(-1)^{J}f^{J-K}_{\alpha}(q)$ for $K<0$, and%
 \beq%
\label{QKK}%
Q^{}_{2\mu}(K^\prime,K_i;q_f,q_i)\equiv\langle\Phi(q_f)\vert
\hat{Q}^{}_{2\mu} \hat{P}^{J_i}_{K^\prime K_i}\vert\Phi(q_i)\rangle.
 \eeq}%

\noindent More details on the calculation of the reduced $E2$ matrix
element are given in Appendix~\ref{Appendix}. The matrix elements of
the charge quadrupole operator $\hat
Q_{2\mu}=e\sum_pr^2_pY_{2\mu}(\Omega_p)$ are calculated in the full
configuration space. There is no need for effective charges, and $e$
simply corresponds to the bare value of the proton charge.

Electric monopole (E0) transitions are calculated from the off-diagonal
matrix elements of the $E0$ operator. The corresponding diagonal matrix
elements are directly related to mean-square charge radii that provide
signatures of shape changes in nuclei. The relation between E0 transitions
and shape transitions and coexistence phenomena has been
extensively investigated \cite{Wood92,Wood99,Zerguine08,Wiedeking08}.
The $E0$ transition rate $\tau(E0)$ between $0^+_1$ and $0^+_2$ can be
separated into two factors: the electronic and the nuclear \cite{Wood99}
\beq
 \dfrac{1}{\tau(E0)} = \rho^2_{21} \Omega \;,
 \eeq
 where the nuclear factor $\rho^2_{21} $ is defined by:
 \beqn
 \rho^2_{21}(E0)
 =\left|\langle 0^+_2\vert \hat T(E0)\vert 0^+_1\rangle\right|^2/e^2R^4_0,
 \eeqn
 and $\hat T(E0)=\sum_ke_kr^2_k$.
 The off-diagonal matrix elements of the
 $E0$ operator can be evaluated using
 angular momentum projected GCM wave functions:
 \beqn
 &&\langle 0^+_2\vert \hat T(E0)\vert 0^+_1\rangle\nonumber\\
 &=&\sum_{q_i,q_j} f^\ast_{0^+_2}(q_j)f_{0^+_1}(q_i)
       \langle\Phi(q_j)\vert\hat T(E0)\hat P^{0}_{00}
       \vert\Phi(q_i)\rangle \;.
 \eeqn
Finally, it will be useful to check the average number of particles
for a collective state $\vert \Psi^{JM}_\alpha\rangle$:
 \begin{eqnarray}
 \label{av_N}
 N^J_\alpha
 &=& \langle \Psi^{JM}_\alpha \vert \hat N\vert \Psi^{JM}_\alpha\rangle\nonumber\\
 &=&\sum_{q_j,q_i;K_1,K_2}\Delta_{K_1K_2}
     f^{\ast JK_2}_{\alpha}(q_j)f^{JK_1}_{\alpha}(q_i)\nonumber\\
 &&\times
     \int d\Omega D^{J\ast}_{K_2K_1}
    \langle\Phi(q_j)\vert\hat N\hat R(\Omega)\vert\Phi(q_i)\rangle \;,
 \end{eqnarray}
where $\hat N=\sum_k a^\dagger_k a_k$ is the particle number operator, and
 \beqn
 \label{PNoverlap}
 \dfrac{\langle\Phi(q_j)\vert\hat N\hat R(\Omega)\vert\Phi(q_i)\rangle}
 {\langle\Phi(q_j)\vert\hat R(\Omega)\vert\Phi(q_i)\rangle}
 =\int d\bm{r} \rho_V(\bm{r};q_j,q_i;\Omega) \;.
 \eeqn
 $\rho_V(\bm{r})$ is the zeroth component of the nucleon
 vector current (cf. Eq.~\ref{dens_3}), and the expression for the
 corresponding transition vector density
 $\rho_V(\bm{r};q_j,q_i;\Omega)$ has been given in
 Ref.~\cite{Yao09amp}. Since the intrinsic state $\vert\Phi(q_i)\rangle$
 corresponds to a BCS wave function, i.e. it is not an eigenstate of the particle number
 operator, the trace of the transition density in Eq.(\ref{PNoverlap}) generally
 does not equal the total nucleon number.

\section{The low-spin spectrum of $^{24}$Mg}
 \label{Sec.III}
In this section we perform several illustrative configuration mixing
calculations that will test our implementation of the 3D angular
momentum projection and the generator coordinate method. The
intrinsic wave functions that are used in the configuration mixing
calculation have been obtained as solutions of the self-consistent
relativistic mean-field equations, subject to constraint on the axial
and triaxial mass quadrupole moments. The interaction in the
particle-hole channel is determined by the relativistic density
functional PC-F1~\cite{Burvenich02}, and a density-independent
$\delta$-force is used as the effective interaction in the
particle-particle channel. Pairing correlations are treated in
the BCS approximation. The pairing strength parameters $V_{\tau}$
($\tau = p,n$) are adjusted by fitting the average gaps of the
mean-field ground state \cite{Dobaczewski84} of $^{24}$Mg
\beq%
\langle\Delta\rangle\equiv%
\frac{\sum_{k}f_{k}v_{k}^{2}\Delta_{k}}{\sum_{k}f_{k}v_{k}^{2}} \;,
\label{avgap}%
\eeq%
to the experimental values obtained from odd-even mass differences
using the five-point formula:  $\Delta^{(5)}_{\rm n}=3.193$ MeV, and
$\Delta^{(5)}_{\rm p}=3.123$ MeV.  The quantities $f_k$ are defined
in Eq.(\ref{Weight}) and $v^2_k$ are the occupation probabilities of
single-nucleon states. The resulting pairing strengths are
$V_n=511.300$ fm$^3$ MeV for neutrons, and $V_p=518.350$ fm$^3$ MeV
for protons. We note that these values differ from the universal
parameters of Ref.~\cite{Burvenich02}, that have been adjusted to
pairing properties of heavy nuclei. With the original pairing
strengths of Ref.~\cite{Burvenich02}, the resulting gaps for
$^{24}$Mg are considerably smaller than the ones obtained from
experimental odd-even mass differences.

Parity, $D_{2}$-symmetry, and time-reversal invariance are imposed in
the mean-field calculation, and this implies that the space-like
components of the single-nucleon four-currents ($j^{\mu},
j^{\mu}_{TV}$) vanish. The scalar ($\rho_{S}, \rho_{TS}$) and vector
($\rho_{V}, \rho_{TV}$) densities in the EDF of Eq.~(\ref{EDF}) are
symmetric under reflections with respect to the $yz$, $xz$ and $xy$
planes. Obviously these symmetries are not fulfilled by the
transition densities and, therefore, the octant $x,y,z \geq 0$ must
be extended to the entire coordinate space when evaluating transition
densities.

To solve the Dirac equation for triaxially deformed potentials, the
single-nucleon spinors are expanded in the basis of eigenfunctions
of a three-dimensional harmonic oscillator (HO) in Cartesian
coordinate~\cite{Koepf88} with $N_{\mathrm{sh}}$ major shells. In
Ref.~\cite{Yao09amp} it has been shown  that $N_{\mathrm{sh}}=8$ is
sufficient to obtain a reasonably converged mean-field potential
energy curve for $^{24}$Mg. The HO basis is chosen isotropic, i.e.
the oscillator parameters $b_{x}=b_{y}=b_{z}=b_{0} = \sqrt{
\hbar/m\omega_{0} } $ in order to keep the basis closed under
rotations~\cite{Egido93,Robledo94}. The oscillator frequency is
given by $\hbar\omega_{0}=41A^{-1/3}$. The Gaussian-Legendre
quadrature is used for integrals over the Euler angles $\phi,\theta$
and $\psi$ in the calculation of the norm and hamiltonian kernels.
With the choice of the number of mesh points for the Euler angles in
the interval $[0,\pi]$: $N_\phi=N_\psi=8$, and $N_\theta=12$, the
calculation achieves an accuracy of $\approx 0.05\%$ for the energy
of a projected state with angular momentum $J \leq 6$ in the
ground-state band~\cite{Yao09amp}.

The nucleus $^{24}$Mg presents an illustrative test case for the
3DAMP+GCM approach to low-energy nuclear structure. The
principal motivation for considering this nucleus is the
direct comparison of the present analysis with the results of
Ref.~\cite{Bender08}, where a 3DAMP+GCM
model has been developed based on Skyrme triaxial mean-field states
that are projected on particle number and angular momentum, and mixed
by the generator coordinate method.
Collective phenomena are, of course, much more pronounced in heavy
nuclei and, therefore, the goal is to eventually apply the present
approach to the rare earth nuclides and the Actinide region.
This will require not only a large oscillator basis, but also a large
number of mesh-points for the Gaussian quadrature in coordinate space,
as well as a finer mesh for the Euler angles and the
deformation parameters.

Axially symmetric AMP+GCM calculations are at present
routinely performed for heavy nuclei~\cite{Niksic06I}, and from such
studies one can estimate that $N_f \approx 16$ shells have to be included in
the oscillator basis for the systems in the mass region around Pb. Note that
the computing time necessary for the evaluation of one overlap matrix element
scales approximately with $N_f^6$. For instance, the
number of mesh-points in the axial deformation $\beta$
that was used in Ref.~\cite{Niksic06I}  is a factor 4 larger than in the present
analysis and,  moreover, in the 3D case one
also needs a finer mesh for the integration over Euler angles.
These considerations show that a straightforward application of the
existing 3DAMP+GCM codes to $A\approx 200$ heavy nuclei
will basically depend on the availability of large-scale
general-purpose computer resources.
On the other hand, the introduction of additional approximations could
considerably reduce the computing requirements. For instance, the overlap
functions are strongly peaked at $q=q'$, and the use of Gaussian
overlap approximations has produced excellent results in many cases.
These approximations form the basis for the derivation of a collective Bohr
Hamiltonian for quadrupole degrees of freedom~\cite{Niksic09,Li09}.

\subsection{Convergence of the 3DAMP+GCM calculations}
\begin{figure}[h]\hspace{-1.5cm}
 \includegraphics[width=10cm]{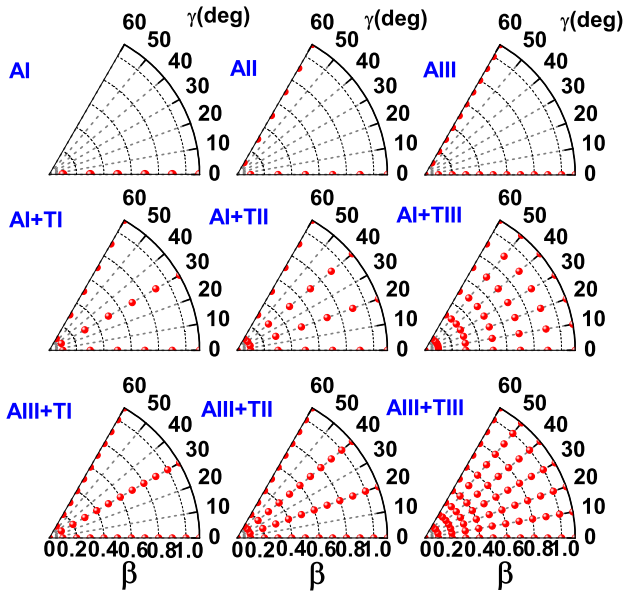}
 \caption{(Color online) The distribution of mesh points in the ($\beta, \gamma$) plane
 for the sets AI, AII, AIII, AI+TI, AI+TII, AI+TIII, AIII+TI, AIII+TII and AIII+TIII.}
 \label{fig2}
\end{figure}

The convergence of the 3DAMP+GCM calculation has been examined with
respect to both the number of mesh points in the ($\beta, \gamma$)
plane, and the cutoff parameter $\zeta$ that is used to remove from
the GCM basis the eigenstates of the norm overlap kernel
$\mathscr{N}^J$ with very small eigenvalues $n^J_k/n^J_{\textrm max}
< \zeta$. In the first step the cutoff is set to
$\zeta=5\times10^{-3}$, and we compare low-lying spectra of $^{24}$Mg
that are obtained in 3DAMP+GCM calculations with different numbers of
points of the discretized generator coordinates. We consider the
following sets of generator coordinates: (AI, AII, AIII) include only axial
deformations (prolate and oblate shapes)
 \begin{itemize}
   \item AI: $(\beta,\gamma)=(0.1, 0^\circ)$, (0.3, 0$^\circ$), (0.5, 0$^\circ$), (0.7, 0$^\circ$),
                   (0.9, 0$^\circ$), (1.1, 0$^\circ$);
   \item AII: $(\beta,\gamma)=(0.1, 0^\circ$), (0.1, 60$^\circ$), (0.3, 0$^\circ$), (0.3, 60$^\circ$), (0.5, 0$^\circ$),
                   (0.5, 60$^\circ$), (0.7, 0$^\circ$), (0.7, 60$^\circ$), (0.9, 0$^\circ$), (0.9, 60$^\circ$), (1.1, 0$^\circ$), (1.1, 60$^\circ$);
   \item AIII: $(\beta,\gamma)=(0, 0^\circ$), (0.1, 0$^\circ$), (0.1, 60$^\circ$),
                   (0.2, 0$^\circ$), (0.2, 60$^\circ$), (0.3, 0$^\circ$), (0.3, 60$^\circ$),  (0.4, 0$^\circ$), (0.4, 60$^\circ$), (0.5, 0$^\circ$),
                   (0.5, 60$^\circ$), (0.6, 0$^\circ$), (0.6, 60$^\circ$), (0.7, 0$^\circ$), (0.7, 60$^\circ$), (0.8, 0$^\circ$), (0.8, 60$^\circ$), (0.9, 0$^\circ$), (0.9, 60$^\circ$),
                   (1.0, 0$^\circ$), (1.0, 60$^\circ$), (1.1, 0$^\circ$), (1.1,
                   60$^\circ$) \;.
 \end{itemize}
 (TI, TII, TIII) denote different sets with $\gamma \neq 0$ (triaxial shapes):
 \begin{itemize}
   \item TI: $\gamma=0^\circ,30^\circ,60^\circ$;
   \item TII: $\gamma=0^\circ,20^\circ,40^\circ,60^\circ$;
   \item TIII: $\gamma=0^\circ,10^\circ,20^\circ,30^\circ,40^\circ,50^\circ,60^\circ$.
 \end{itemize}
The sets of ($\beta, \gamma$) mesh-points
shown in Fig.~\ref{fig2} have been used in the present analysis.

  \begin{table}[]
   \centering
   \tabcolsep=8pt
   \caption{Ground state energies $E_{gs}$, excitation energies $E_x$ (in MeV) and $B(E2)$ values (in e$^2$fm$^4$)
   for transitions between low-spin states in $^{24}$Mg, calculated with the
   3DAMP+GCM model for the generator coordinate sets AI, AII and AIII
   (see text for details). }
   \begin{tabular}{lrrrrr}
    \hline\hline
    quantities                    &  AI      &  AII       &  AIII          \\ \hline
   $E_{gs}$($0^+_1$)              &  -196.985&  -197.291  & -197.279       \\
   $E_x$($2^+_1$)                 & 2.196    & 2.351      &    2.330        \\
   $E_x$($4^+_1$)                 & 5.394    & 5.905      &    5.849       \\
   $E_x$($6^+_1$)                 & 10.426   & 10.591     &    10.568      \\
   $B(E2: 2^+_1\rightarrow0^+_1$) &  78.155  & 78.721     &    79.135      \\
   $B(E2: 4^+_1\rightarrow2^+_1$) &  137.679 &140.814     &   139.750      \\
   $B(E2: 6^+_1\rightarrow4^+_1$) &  177.025 &169.246     &   168.527      \\
    \hline\hline
 \end{tabular}
 \label{tab1}
 \end{table}

  \begin{table}[]
   \centering
   \tabcolsep=8pt
   \caption{Same as Table~\ref{tab1}, but for the
   generator coordinates sets AI+TI, AI+TII and AI+TIII (see text for details).}
   \begin{tabular}{lrrrrrrr}
    \hline\hline
   quantities                    &  AI+TI   &  AI+TII   &  AI+TIII          \\
   \hline
   $E_{gs}$($0^+_1$)             & -197.285 &-197.304  & -197.307               \\
   $E_x$($2^+_1$)                &  2.241   &2.198     & 2.177             \\
   $E_x$($4^+_1$)                &  5.776   &5.725     & 5.677              \\
   $E_x$($6^+_1$)                &  10.485  &10.413    & 10.360           \\
   $B(E2: 2^+_1\rightarrow0^+_1$)&  80.523  &80.849    &  81.435           \\
   $B(E2: 4^+_1\rightarrow2^+_1$)& 144.441  &145.926   & 147.178             \\
   $B(E2: 6^+_1\rightarrow4^+_1$)& 171.275  &178.015   & 182.199             \\
    \hline\hline
 \end{tabular}
 \label{tab2}
 \end{table}
In Table~\ref{tab1} we display the excitation energies and $B(E2)$
values of low-spin yrast states in $^{24}$Mg, calculated with the
3DAMP+GCM model, but including only axially deformed mean-field
states (coordinate sets AI, AII and AIII, as shown in
Fig.~\ref{fig2}). One notes that the largest differences in the
calculated excitation energies are within 10 \% and the
$B(E2:J\rightarrow J-2)$ values agree within 5 \%. The major step
for the energies comes from the inclusion of oblate shapes (AII) in
the GCM configuration mixing calculations. It lowers the total
ground state energy by $\approx 300$ keV and increases the
energies by $\approx 150$ keV for the $2^+_1$, 200 keV
for the $4^+_1$, and 150 keV for the $6^+_1$ state. The refinement
of the mesh in AIII produces only small changes.

Similar results are found when comparing results of 3DAMP+GCM
calculations based on triaxial intrinsic states: AI+TI, AI+TII and
AI+TIII in Table~II, and AIII+TI, AIII+TII and AIII+TIII in
Table~III. The effect of including triaxial deformations, i.e. the
$\gamma$ degree of freedom, is perhaps best illustrated in the
comparison between results obtained with the sets of generator
coordinates AIII (Tab.~\ref{tab1}) and AIII+TIII (Tab.~\ref{tab3}).
The inclusion of triaxial states in the GCM configuration mixing
calculation lowers the total energies by 39 keV for $0^+_1$, 180
keV for $2^+_1$, 226 keV for $4^+_1$, and 262 keV for $6^+_1$.
The corresponding $B(E2:J\rightarrow J-2)$ values are enhanced by
$3.94\%, 5.29\%, 7.63\%$ for $J^\pi_\alpha=2^+_1, 4^+_1, 6^+_1$,
respectively.

The influence of the $\gamma$ degree of freedom, and the convergence
of 3DAMP+GCM calculations with respect to the number of mesh points
of the discretized generator coordinates can clearly be seen in the
comparison of calculations with mean-field states at the mesh points
of coordinate sets AI, AI+TIII and AIII, AIII+TIII. The inclusion of
triaxial shapes lowers the energies by $\approx 300$ keV.
On the other hand, very similar results are obtained in calculations
based on coordinate sets that differ only in the number of axial
points. Therefore, we find that, if prolate as well as oblate
configurations are included, the spectroscopic properties of low-spin
states in $^{24}$Mg are not very sensitive to the number of axial
meshpoints. The inclusion of the $\gamma$ degree of freedom changes
this situation somewhat, but not dramatically for the ground stated
band where the admixtures with $K\neq 0$ are small. This is
consistent with the results of the 3DAMP+GCM calculation with
particle-number projection \cite{Bender08}, based on the
non-relativistic Skyrme density functional. It was shown, namely,
that the number of axial states that can be added to the set of
triaxial states is not large. Redundancies appear very quickly in the
norm kernel when more states are added to the nonorthogonal basis,
and this is simply a consequence of very few level crossings as
function of deformation in $^{24}$Mg.
\begin{table}[]
   \centering
   \tabcolsep=8pt
   \caption{Same as Table~\ref{tab1}, but for the
   generator coordinates sets AIII+TI, AIII+TII and AIII+TIII (see text for details).}
   \begin{tabular}{lrrrrrrr}
   \hline\hline
   quantities                    &   AIII+TI   &  AIII+TII &  AIII+TIII        \\
   \hline
   $E_{gs}$($0^+_1$)             &  -197.290   &  -197.306  &   -197.318          \\
   $E_x$($2^+_1$)                &  2.239      & 2.205       &   2.189                  \\
   $E_x$($4^+_1$)                &  5.735      & 5.695       &   5.662                 \\
   $E_x$($6^+_1$)                &  10.452     & 10.388      &   10.345             \\
   $B(E2: 2^+_1\rightarrow0^+_1$)&  80.498     &  81.488     &    82.256           \\
   $B(E2: 4^+_1\rightarrow2^+_1$)& 143.042     & 145.525     &   147.137             \\
   $B(E2: 6^+_1\rightarrow4^+_1$)& 166.952     &  175.157    &   181.379               \\
    \hline\hline
 \end{tabular}
 \label{tab3}
 \end{table}
In Table~\ref{tab5} we show the excitation energies and $B(E2)$
values for low-lying states in $^{24}$Mg, calculated with the
3DAMP+GCM model based on a set of axial mean-field states with
$\beta=0, 0.1, 0.2, \cdots, 1.1$ and $\gamma=0^\circ, 60^\circ$, as
functions of the cutoff parameter $\zeta$, that defines the basis of
``{\em natural states}".  Eigenstates of the norm overlap kernel
$\mathscr{N}^J$ with eigenvalues $n^J_k/n^J_{\textrm max} < \zeta$
are removed from the GCM basis ($n^J_{\textrm max}$ is the largest
eigenvalue of the norm kernel for a given angular momentum). The
excitation energies are not sensitive to the particular value of the
cutoff parameter provided $\zeta<1\times10^{-2}$, whereas the effect
on the $B(E2)$ values is $ < 1\%$ for smaller values of $\zeta$.
However, $\zeta$ cannot be taken arbitrarily small, because spurious
states are introduced in the basis for very small eigenvalues of the
norm overlap kernel. The remaining calculations presented in this
work have been performed using the value $\zeta=5\times10^{-3}$.

\begin{table*}[]
   \centering
   \tabcolsep=10pt
   \caption{Excitation energies $E_x$ (in MeV) and $B(E2)$ values (in e$^2$fm$^4$)
   for transitions between low-spin states in $^{24}$Mg, calculated with the
   3DAMP+GCM model for the generator coordinates $\beta=0, 0.1, \cdots, 1.1$,
   and $\gamma=0^\circ, 60^\circ$ as functions of the cutoff parameter $\zeta$ that defines
   the basis of ``{\em natural states}".}
   \begin{tabular}{lrrrrrrrr}
    \hline\hline
   $\zeta$                       &   $5\times10^{-2}$  &  $1\times10^{-2}$   & $5\times10^{-3}$       & $1\times10^{-3}$       & $5\times10^{-4}$  & $1\times10^{-4}$      \\ \hline
   $E_x$($2^+_1$)                &     2.275           &     2.340           &    2.330               &   2.341                &   2.314           &  2.320                  \\
   $E_x$($4^+_1$)                &     5.703           &     5.931           &    5.849               &   5.573                &   5.544           &  5.580                 \\
   $B(E2: 2^+_1\rightarrow0^+_1$)&    77.586           &    80.138           &   79.135               &  79.967                &  79.113           &  79.688          \\
   $B(E2: 4^+_1\rightarrow2^+_1$)&   144.403           &   143.314           &  139.750               & 136.372                & 137.989           & 138.669                 \\
    \hline\hline
 \end{tabular}
 \label{tab5}
 \end{table*}

 \begin{figure}[h!]
 \includegraphics[width=6cm]{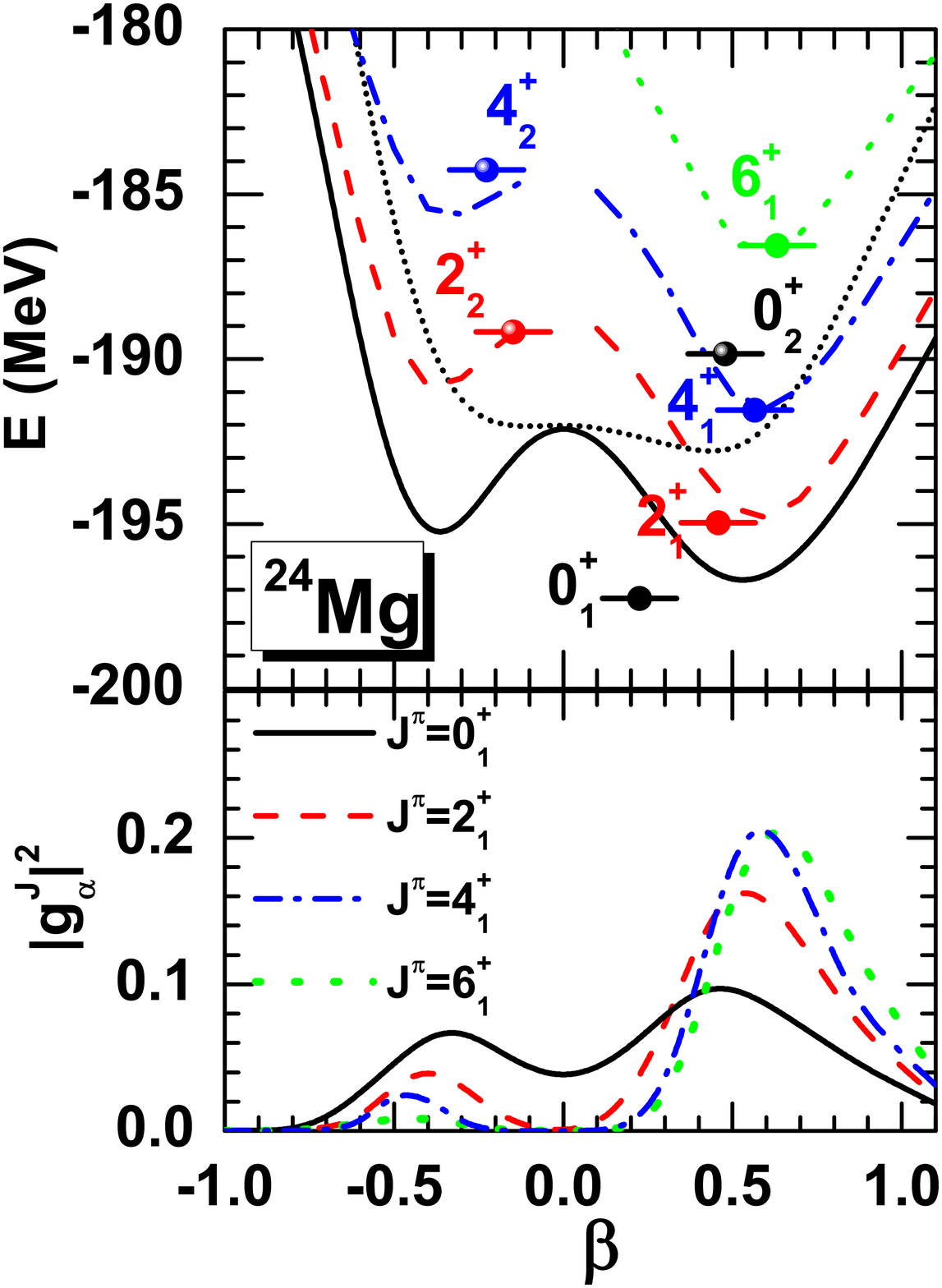}
 \caption{(Color online)  Upper panel: energies and the average axial deformations for
  the two lowest GCM states with angular momentum $0^+, 2^+, 4^+, 6^+$ in $^{24}$Mg,
  together with the mean-field (dotted) and the corresponding angular-momentum
  projected energy curves.  Lower
  panel: squares of collective wave functions $\vert g^J_\alpha(q)\vert^2$
  with $q_{22}=0$ for the corresponding lowest GCM states in $^{24}$Mg.
  These results are obtained in the axial 1DAMP+GCM calculation. Positive (negative)
  values of the axial deformation $\beta$ correspond to prolate (oblate) configurations.}
 \label{fig3}
 \end{figure}

 \subsection{Axially-symmetric AMP+GCM calculation}

 \begin{figure}[]
 \centering
 \includegraphics[width=9cm]{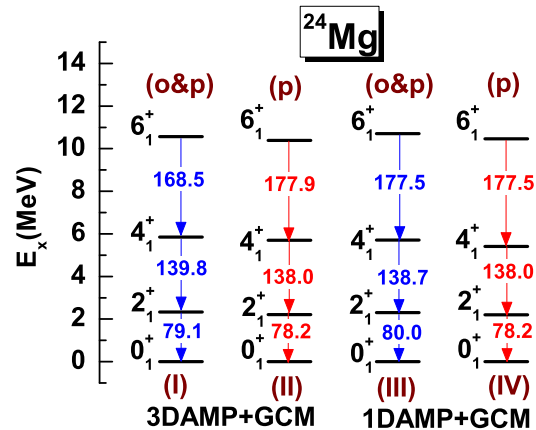} 
 \caption{(Color online) Lowest energy levels of angular
 momentum $J^\pi = 0^+$, $2^+$, $4^+$, $6^+$ in $^{24}$Mg,
 and reduced $E2$ transition probabilities in e$^2$fm$^4$,
 calculated with the
 3DAMP+GCM and 1DAMP+GCM models. See text for details.}
 \label{fig4}
 \end{figure}

By restricting the set of intrinsic states to axially symmetric
configurations: $\gamma=0$ and $\gamma=180^\circ$, the complicated
3DAMP+GCM model is reduced to a relatively simple 1DAMP+GCM
calculation. For the choice of generator coordinates $\beta=0, 0.1,
0.2, \cdots ,1.1$; $\gamma=0$ and $\gamma=180^\circ$, we have
calculated the energies and the average axial quadrupole deformations
of the two lowest GCM states, for each angular momentum: $0^+$,
$2^+$, $4^+$, and $6^+$ in $^{24}$Mg, as shown in Fig.~\ref{fig3}.

The mean-field  energy surface is somewhat soft with a prolate deformed
minimum at $\beta \approx 0.50, \gamma=0^\circ$, and the total energy
$E=-192.807$ MeV. This result is consistent with our previous
calculation that used the PC-F1 energy density functional plus a
monopole pairing force~\cite{Yao08cpl}, and with an earlier study that
employed the relativistic mean-field model with the NL2 effective
interaction~\cite{Koepf88}. A rotational yrast band is calculated in
the prolate minimum,  with the squares of collective wave functions
(probabilities) concentrated at $\beta \approx 0.5$.

In Fig.~\ref{fig4} we display the lowest energy levels of angular
momentum $J^\pi = 0^+$, $2^+$, $4^+$, $6^+$ in $^{24}$Mg, calculated
with the 3DAMP+GCM and 1DAMP+GCM codes, for the sets of axially
symmetric generator coordinates: $\beta=0, 0.1, 0.2, \cdots ,1.1$
with both prolate ($\gamma=0$) and oblate states ($\gamma=60^\circ$
and $\gamma=180^\circ$ in 3DAMP+GCM and 1DAMP+GCM models,
respectively) (columns I and III), and with only prolate states
$\gamma=0$ (columns II and IV). As expected, the 3DAMP+GCM and
1DAMP+GCM calculations produce virtually identical results, with small
differences attributed to the
numerical accuracy. In fact, the difference between the  $B(E2)$
values shown in columns I and III can be further reduced by increasing
the number of mesh-points used in the Gaussian-Legendre quadrature
over the Euler angles $\phi,\theta$ and $\psi$ in the calculation of the norm
and hamiltonian kernels.

 \subsection{Triaxial AMP+GCM calculation}
\begin{figure}[h!]
 \includegraphics[width=9cm]{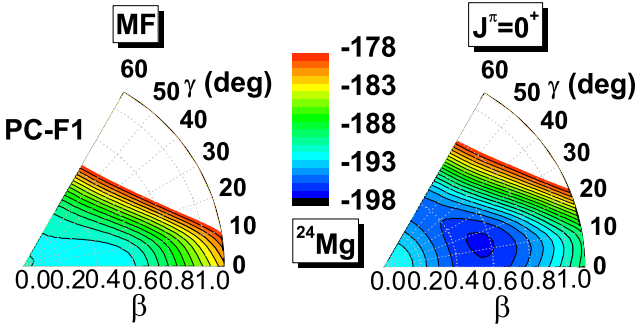}
 \caption{(Color online) Self-consistent RMF+BCS energy surface
 (left panel) of $^{24}$Mg in the $\beta$-$\gamma$
 plane ($0\le \gamma\le 60^0$), and angular momentum projected energy surface with
 $J^\pi=0^+$  (right panel). The contours join points on the surface
 with the same energy. The difference between neighboring contours is
 1.0 MeV.}
 \label{fig5}
 \end{figure}
 \begin{figure}[h!]
 \includegraphics[width=8cm]{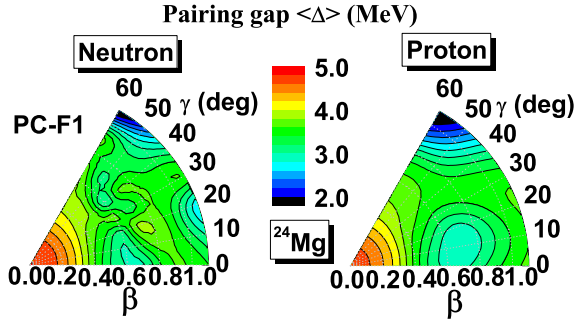}
 \caption{(Color online) Average neutron and proton pairing gaps of $^{24}$Mg
in the $\beta$-$\gamma$ plane ($0\le \gamma\le 60^0$).  The contours
join points on the surface with the same pairing gap. The difference
between neighboring contours is 0.2 MeV.}
 \label{fig6}
 \end{figure}

 \begin{figure*}[]
 \centering
 \includegraphics[width=18cm]{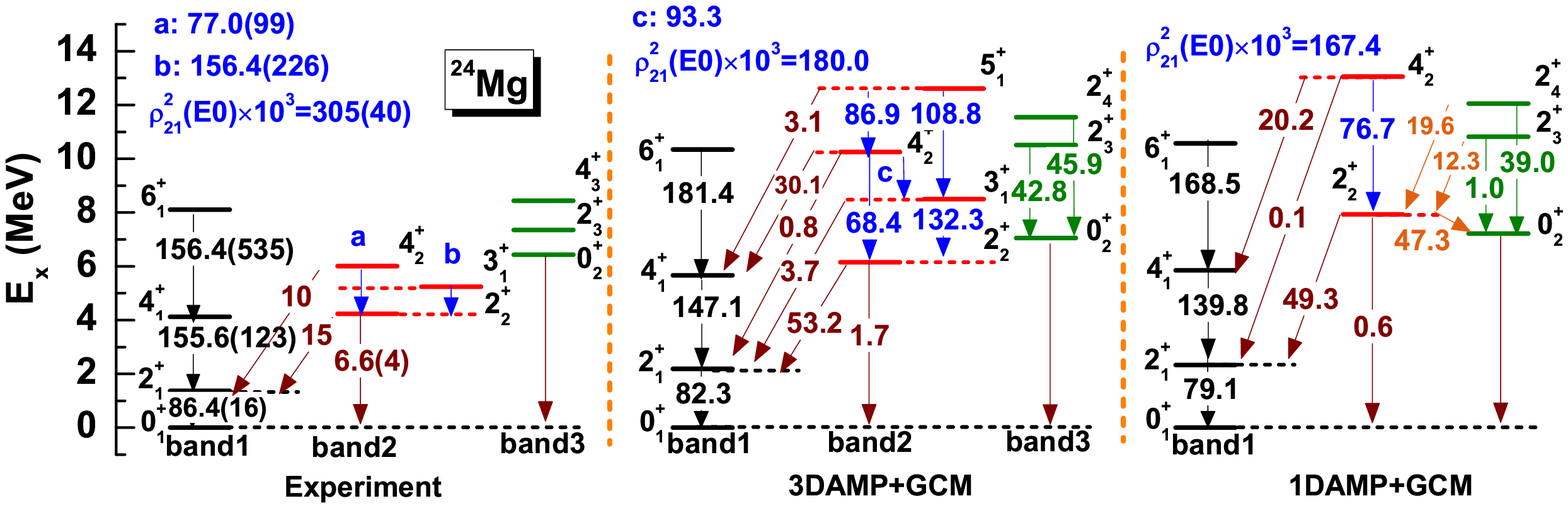} 
 \caption{(Color online) The low-spin level scheme of $^{24}$Mg
 calculated using the 3DAMP+GCM model and 1DAMP+GCM model with the PC-F1
 relativistic density functional, in comparison with
 data \cite{Endt93,Branford75,Keinonen89}.
 The $B(E2)$ values are given in units of e$^2$fm$^4$.}
 \label{fig7}
 \end{figure*}

In Fig.~\ref{fig5} we plot the self-consistent RMF+BCS triaxial
energy surface of $^{24}$Mg in the $\beta$-$\gamma$ plane ($0\le
\gamma\le 60^0$), obtained by imposing constraints on the expectation
values of the quadrupole moments $q_{20}$ and ${q}_{22}$. The panel
on the right displays the projected energy surface with $J^\pi=0^+$:
\begin{equation}
E^{J=0} (q) = {{\mathscr{H}^{J=0}(q,q)} \over {\mathscr{N}^{J=0}(q,q)}}
\end{equation}
The contours join points with the same energy and the difference
between neighboring contours is 1.0 MeV. The energy surfaces nicely
illustrate the effects of including triaxial shapes and of the
restoration of rotational symmetry. The mean-field energy surfaces
are found to be quite soft with a minimum at an axial prolate
deformation $\beta \approx 0.5$. When compared with the axial plot in
Fig.~\ref{fig3}, one realizes that the oblate minimum on the axial
projected energy curve with $J^\pi=0^+$ is actually a saddle point in
$\gamma$ direction. Projection shifts the minimum to a slightly
triaxial shape with $\beta=0.50, \gamma=20^\circ$ and $E=-197.074$
MeV. The gain in energy from the restoration of rotational symmetry
is $4.266$ MeV. The fact that angular momentum projection leads to
triaxial minima in the PES was already noted in 3DAMP
calculations in the eighties~\cite{Hayashi84}, and very similar
results have been obtained recently~\cite{Bender08} for the
nucleus $^{24}$Mg using the Skyrme functional SLy4. We note, however,
that the 3DAMP+GCM model used in Ref.~\cite{Bender08} includes a
projection on proton and neutron numbers, that is not carried out in
the present analysis.

Fig.~\ref{fig6} displays the corresponding average neutron and
proton pairing gaps $\langle\Delta\rangle$, defined by
Eq.~(\ref{avgap}), as functions of deformation variables $\beta$ and
$\gamma$. The gaps are relatively small around the minimum of the
potential energy surface (PES), whereas larger values are calculated
at the saddle points. The fluctuations of pairing gaps reflect the
underlying shell structure.

 \begin{figure}[]
 \centering
 \includegraphics[width=9cm]{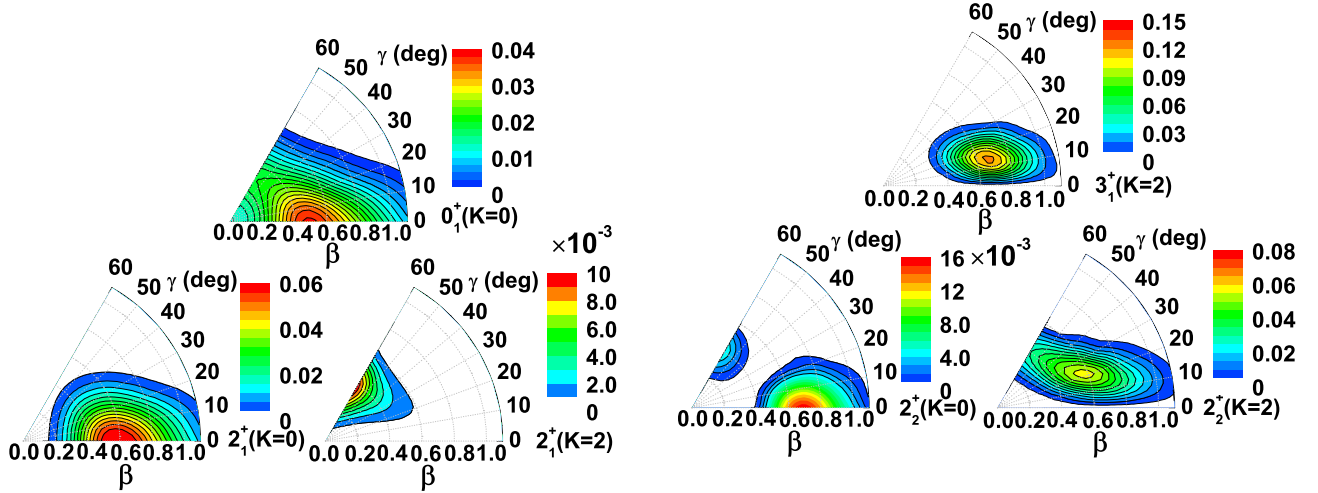}
 \caption{(Color online) Contour plots of the probability distributions $\vert g^J_\alpha\vert^2$
 for the ground state $0^+_1$
 and the first excited state $2^+_1$ (both the $K=0$ and $K=2$ components) in $^{24}$Mg. }
 \label{fig8}
\end{figure}

\begin{figure}[]
 \centering
 \vspace{-0.5cm}
 \includegraphics[width=9cm]{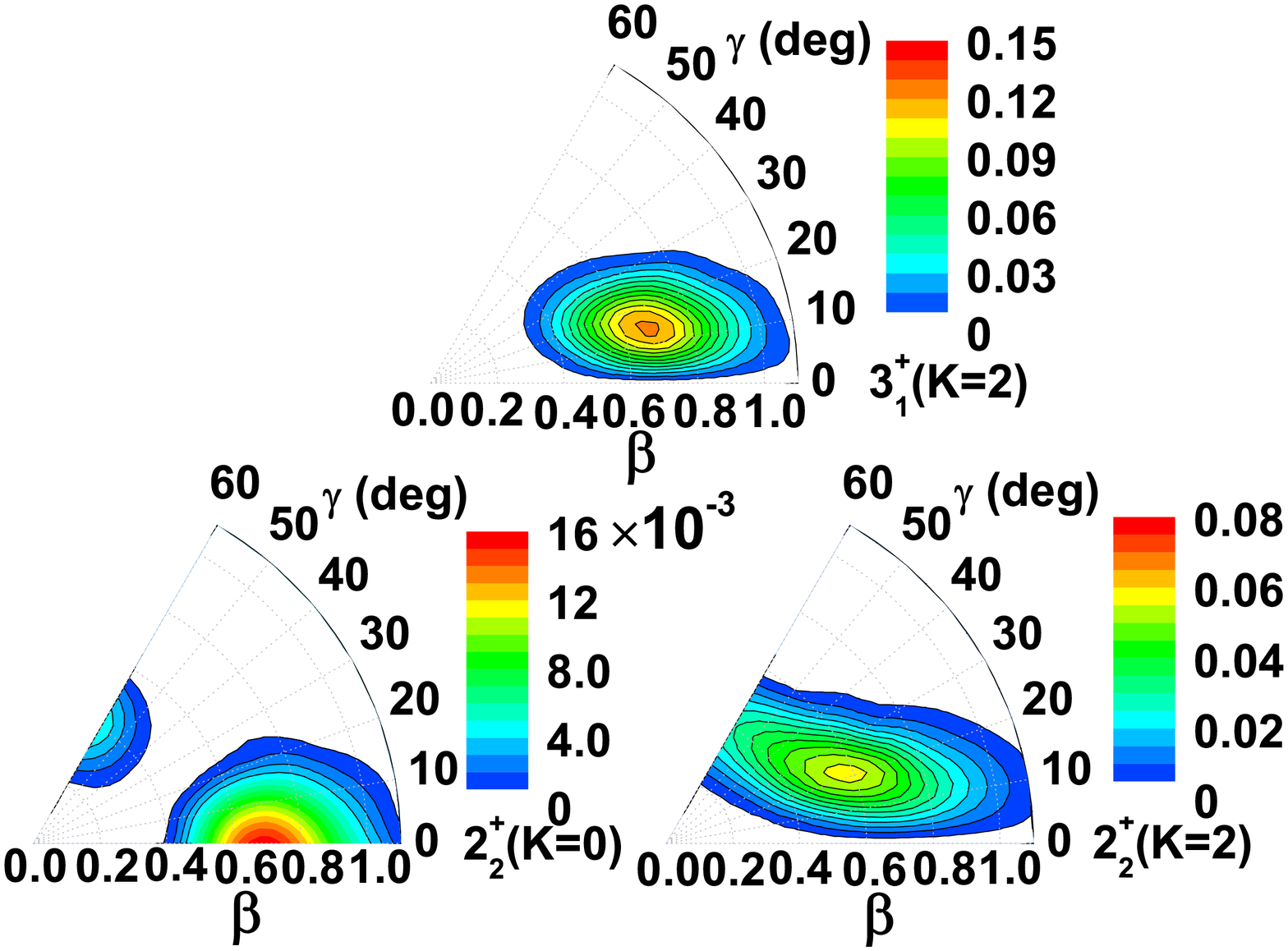}
 \caption{(Color online) Same as in Fig.~\ref{fig8}, but  for the
 excited states $3^+_1$ and $2^+_2$ in $^{24}$Mg.}
 \label{fig9}
\end{figure}

\begin{figure}[]
 \centering
 \vspace{-0.5cm}
 \includegraphics[width=9cm]{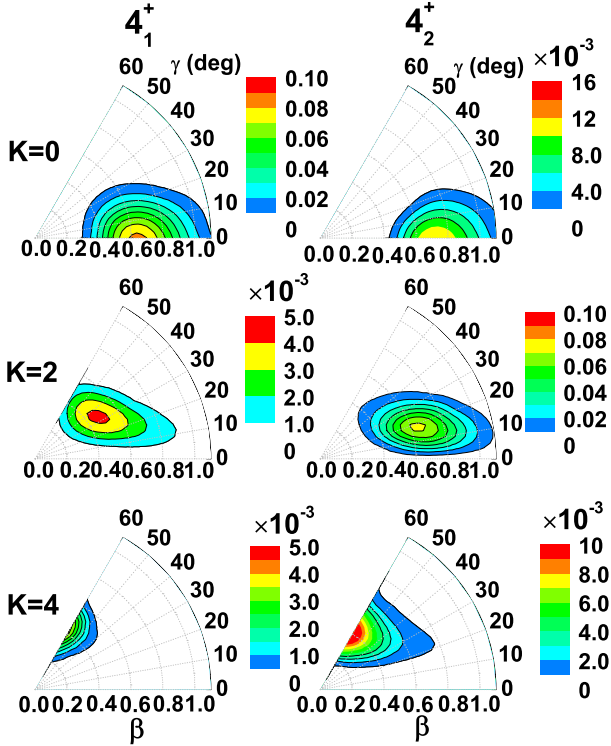}
 \caption{(Color online) Same as in Fig.~\ref{fig8}, but  for the
 excited states $4^+_1$ and $4^+_2$ in $^{24}$Mg.}
 \label{fig10}
\end{figure}

The solution of the HWG equation (\ref{HWEq}) yields the excitation
energies and the collective wave functions for each value of the
total angular momentum and parity $J^\pi$. In addition to the yrast
ground-state band, in deformed and transitional nuclei excited
states are usually also assigned to (quasi) $\beta$ and $\gamma$
bands. This is done according to the distribution of the angular
momentum projection $K$ quantum number
(Figs.~\ref{fig8}-\ref{fig10}). Excited states with predominant
$K=2$ components in the wave function are assigned to the
$\gamma$-band, whereas the $\beta$-band comprises states above the
yrast characterized by dominant $K=0$ components. As an example, in
Fig.~\ref{fig7} we display the low-spin PC-F1 excitation spectrum of
$^{24}$Mg obtained by the 1DAMP+GCM calculation with the AIII set of
generator coordinates, and by the 3DAMP+GCM calculation with the
AIII+TIII set of mesh points, in comparison with available data
\cite{Endt93,Branford75,Keinonen89}. The level scheme is in rather
good agreement with data, but in both cases the calculated spectra
are systematically stretched as compared to experimental bands. This
is because angular-momentum projection is performed only after
variation and, therefore, time-odd components and alignment effects
are neglected. Cranking calculations, for instance, correspond to an
approximate angular-momentum projection before
variation~\cite{Beck70}, and lead to an enhancement of the moments
of inertia in better agreement with data
\cite{Koepf89,Afanasjev00c}. However, at present the full 3D
angular-momentum projection before variation, plus GCM configuration
mixing, is still beyond the available computing capacities. The
agreement of the calculated quadrupole transition probabilities with
data in Fig.~\ref{fig7} is remarkable, especially considering that
the calculation of B(E2) values is parameter-free, i.e. the
transitions are calculated employing bare proton charges.

 \begin{figure}[]
 \centering
 \includegraphics[width=9cm]{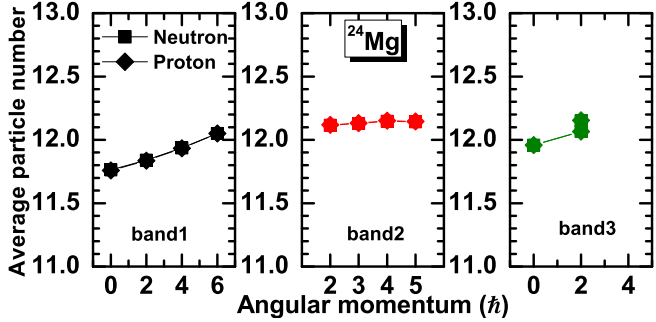} 
 \caption{(Color online) Average particle numbers
 for the 3DAMP+GCM states belonging to the three bands of $^{24}$Mg in
  Fig.~\ref{fig7}.}
 \label{fig11}
 \end{figure}

In Fig.~\ref{fig8}, we plot the corresponding distributions $\vert
g^J_\alpha\vert^2$ of Eq.~(\ref{probability}), with respect to
$\beta$ and $\gamma$, for the ground state $0^+_1$ and the first
excited state $2^+_1$ (both the $K=0$ and $K=2$ components) in
$^{24}$Mg. These quantities give the probabilities that the intrinsic
wave functions of the corresponding states have a certain quadrupole
deformation characterized by the collective coordinates
$\beta$ and $\gamma$. For the ground
state, and for the $K=0$ component of $2^+_1$, these
distributions are largely concentrated along the prolate symmetry
axis. Since the $K = 0$ component of the $2^+_1$ state exhausts
$92\%$ of the norm, this state obviously belongs to the $K=0$ band
built on the nearly prolate ground state. From the PES shown in the right panel
of Fig.~\ref{fig5}, with the pronounced minimum at $\gamma \approx 20^0$,
one would have expected the maximum of
the probability distributions in this region of the ($\beta, \gamma$) plane.
However, it turns out that the inclusion of quadrupole fluctuations through GCM
configuration mixing, drives the structure built on the ground state
back toward the prolate symmetry axis, i.e. the GCM model calculation
does not predict the existence of a stable triaxial structure of the intrinsic
states of the ground-state band of $^{24}$Mg. The probability distributions for
the excited states $3^+_1$ and $2^+_2$ are shown in Fig.~\ref{fig9}.
For $2^+_2$ state the $K = 2$ component exhausts about $87\%$ of the
norm and, therefore, $2^+_2$ and $3^+_1$ are assigned to the $K=2$
(quasi) $\gamma$ band. The $K = 0$, $K = 2$ and $K = 4$ probability
distributions of the states $4^+_1$ and $4^+_2$ are displayed in
Fig.~\ref{fig10}. Since the $K = 0$ ($K = 2$) component of the state
$4^+_1$ ($4^+_2$) exhausts $92\%$ ($79\%$) of the norm, $4^+_1$
belongs to the ground-state band, and $4^+_2$ to the (quasi) $\gamma$
band.

Finally in Fig.~\ref{fig11} we display the average neutron and
proton numbers (cf. Eq.~(\ref{av_N})) for the 3DAMP+GCM states
belonging to the three bands of $^{24}$Mg in Fig.~\ref{fig7}. The
dispersion of the particle number is relative large $( \approx 0.3)$
for states $0^+_1$ and $2^+_1$.

\section{Summary and Outlook}
 \label{Sec.IV}

The framework of relativistic energy density functionals has been
very successfully applied to the description of a rich variety of
structure phenomena over the whole nuclear chart. However, to go
beyond the modeling of bulk nuclear properties and perform detailed
calculations of excitation spectra and transition probabilities, one
must extend the simple single-reference (mean-field) implementation
of this framework, and include long-range correlations related to
restoration of symmetries broken by the static mean field and to
fluctuations of collective coordinates around the mean-field
minimum. Building on recent models \cite{Niksic06I,Niksic06II} that
have employed the generator coordinate method (GCM) to perform
configuration mixing of axially-symmetric relativistic mean-field
wave functions, and especially on Ref.~\cite{Yao09amp}, where we
have already considered three-dimensional angular-momentum
projection (3DAMP) of relativistic mean-field wave functions, in
this work a model has been developed that uses the GCM in
configuration mixing calculations that involve 3DAMP wave functions,
generated  by constrained self-consistent mean-field calculations
for triaxial nuclear shapes.

The current implementation of the relativistic 3DAMP+GCM model has
been tested in the calculation of spectroscopic properties of
low-spin states in $^{24}$Mg. Starting with the relativistic density
functional PC-F1~\cite{Burvenich02}, and a density-independent
$\delta$-force as the effective interaction in the pairing channel,
the intrinsic wave functions are generated from the self-consistent
solutions of the constrained RMF+BCS equations in the basis of a
three-dimensional harmonic oscillator in Cartesian coordinates. The
constraints are on the axial and triaxial mass quadrupole moments.
After restoring rotational symmetry by 3DAMP, the fluctuations of
quadrupole deformations are included by performing GCM mixing of
angular-momentum projected configurations that correspond to
different values of the generator coordinates $\beta$ and $\gamma$.
The GCM calculation has been tested both with respect to the number
of mesh-points in the discretized ($\beta, \gamma$) plane, and the
cutoff-parameter that is used to eliminate from the GCM basis the
``high momentum" eigenvectors of the norm overlap kernels with
extremely small eigenvalues. Results for excitation energies in the
ground-state, (quasi) $\gamma$ and $\beta$ bands, and the
corresponding interband and intraband transition probabilities have
been compared with available data on low-spin states in  $^{24}$Mg.
The comparison has shown a very good agreement between data and the
predictions of the relativistic 3DAMP+GCM model.

The choice of $^{24}$Mg allows a direct comparison of the present
analysis with the results of Ref.~\cite{Bender08}, where a 3DAMP+GCM
model has been developed based on Skyrme triaxial mean-field states
that are projected on particle number and angular momentum, and mixed
by the generator coordinate method. Because it includes projection on
particle number, the model of Ref.~\cite{Bender08} is much more
involved and the numerical implementation is more difficult. In
particular, the use of general EDFs in GCM calculations, i.e.
energy functionals with an arbitrary dependence on nucleon densities,
leads to discontinuities or even divergences of the energy kernels as
functions of deformation, that can possibly produce spurious contaminations
in the calculated excitation spectra (for a detailed discussion, we refer the
reader to Refs.~\cite{Anguiano01,Bender09,Lacroix08}, and references
cited therein). Even though the
results of the present calculation for $^{24}$Mg are in good
agreement with those of Ref.~\cite{Bender08}, an important advantage
of performing particle-number projection is that it prevents a
collapse of pairing when the level density around the Fermi energy is
reduced as, for instance, close to the minimum of the potential
energy surface. The comparison with Ref.~\cite{Bender08} thus points
to an obvious improvement of our 3DAMP+GCM model, i.e. the
implementation of particle-number projection.

As an alternative approach to five-dimensional quadrupole dynamics
that includes rotational symmetry restoration and takes into account
triaxial quadrupole fluctuations, one can construct a collective Bohr
Hamiltonian with deformation-dependent parameters. In a recent
work~\cite{Niksic09}, we have developed a new implementation for the
solution of the eigenvalue problem of a five-dimensional collective
Hamiltonian for quadrupole vibrational and rotational degrees of
freedom, with parameters determined by constrained self-consistent
relativistic mean-field calculations for triaxial shapes. As in the
present work, in addition to the self-consistent mean-field potential
of the PC-F1 relativistic density functional in the particle-hole
channel, for open-shell nuclei pairing correlations are included in
the BCS approximation. In Ref.~\cite{Li09}, the model has been
applied in the study of shape phase transitions in the region $Z =
60$, $62$, $64$ with $N \approx 90$. The collective Hamiltonian can
be derived in the Gaussian overlap approximation (GOA)~\cite{Ring80}
to the full five-dimensional GCM. With the assumption that the GCM
overlap kernels can be approximated by Gaussian functions, the local
expansion of the kernels up to second order in the non-locality
transforms the HWG equation into a second-order differential equation
for the collective Hamiltonian. Therefore, having developed both the
five-dimensional quadrupole collective Hamiltonian, and the full
3DAMP+GCM model, we plan to perform microscopic tests of the GOA in a
study of low-spin spectroscopy of $\gamma$-soft transitional nuclei,
especially the effect of GOA on the calculated transitions between
bands. In general, we expect that both models will be a useful
addition to the theoretical tools that can be used in studies of
complex structure phenomena in medium-heavy and heavy nuclei,
including exotic systems far from stability.


 \begin{acknowledgments}
This work was partly supported by the Asia-Europe Link Project
[CN/ASIA-LINK/008 (094-791)] of the European Commission, Major State
973 Program 2007CB815000 and the National Natural Science Foundation
of China under Grant Nos. 10947013, 10975008 and 10775004, the
Southwest University Initial Research Foundation Grant to Doctor
(No. SWU109011), the DFG cluster of excellence \textquotedblleft
Origin and Structure of the Universe\textquotedblright\
(www.universe-cluster.de), by MZOS - project 1191005-1010, and by
the Chinese-Croatian project "Nuclear structure far from stability".
 \end{acknowledgments}

\begin{appendix}

 \section{Reduced matrix element of the quadrupole operator}
\label{Appendix}
 The basic expressions for the calculation of EM
transition probabilities in the framework of an AMP+GCM approach are
given in Ref.~\cite{Guzman02npa}. Here we start from the reduced
matrix element of the quadrupole operator in
Eq.~(\ref{Integration2}) and derive a formula for the overlap matrix
elements Eq.~(\ref{QKK}):
\beqn%
&&Q_{2\mu}(K^\prime,K;q_i,q_j) \equiv \langle\Phi(q_i)\vert \hat
Q_{2\mu} \hat{P}^{J}_{K^\prime K}\vert\Phi(q_j)\rangle~~~~~~\nonumber\\
&&~~~~~~~=~\frac{2J+1}{8\pi^2}\int d\Omega D^{J\ast}_{K^\prime
K}(\Omega) \langle\hat{Q}_{2\mu} \hat R(\Omega) \rangle_{ij},
\eeqn%
with the overlap function of the quadrupole operator%
\beqn%
\langle\hat{Q}_{2\mu} \hat R(\Omega)
\rangle_{ij}&\equiv&\langle\Phi(q_i)\vert\hat{Q}_{2\mu}
\hat R(\Omega)\vert\Phi(q_j) \rangle\nonumber\\
 &=& \textrm{Tr}[ Q_{2\mu}\rho^{ij}(\Omega)]
 \,\langle \hat R(\Omega) \rangle_{ij}.
\eeqn%
The expressions for the norm overlap $\langle \hat R(\Omega)
\rangle_{ij}$ and transition densities $\rho^{ij}(\Omega)$ are
given in Eqs.~(A28) and (C4) of Ref.~\cite{Yao09amp}.

The indices $q_i, q_j$ run over all  generator coordinates. For
$n_q$ points on the coordinate mesh, only $n_q(n_q+1)/2$ overlaps
need to be evaluated, for instance those with $q_i\leq q_j$. The
remaining part with $q_i>q_j$ is determined by simply exchanging the
indices $i$ and $j$:
\beqn%
 &&Q_{2\mu}(K^\prime,K;q_j,q_i)~~~~~~~~~~~~~~~~~~\\
 &&~~~=~\frac{2J+1}{8\pi^2}\sum_{\mu^\prime}
 \int d\Omega D^{J}_{ KK^\prime}(\Omega)  D^{2}_{\mu^\prime\mu}(\Omega)
  \langle \hat{Q}^\dagger_{2\mu^\prime}\hat R(\Omega)
  \rangle^\ast_{ij}.
\nonumber%
\eeqn%
In the derivation of above relation, an irreducible tensor $\overline
Q_{2-\mu}$
 has been introduced as $Q^\dagger_{2\mu}=(-1)^{\mu}\overline
 Q_{2-\mu}$.

The matrix elements of the multipole moment operator
$\hat{Q}_{\lambda\mu}=r^\lambda Y_{\lambda\mu}$ in the spherical
harmonic oscillator basis $\vert nljm\rangle$ read:
\beqn%
 (Q_{\lambda\mu})_{mm^\prime}
 =\langle nl\vert r^\lambda\vert  n^\prime l^\prime\rangle\cdot
 \langle ljm\vert Y_{\lambda\mu}\vert l^\prime j^\prime m^\prime\rangle \;.
\eeqn%
The radial part is given by
\beqn%
&&\langle nl\vert r^\lambda\vert  n^\prime l^\prime\rangle=%
\frac{(-1)^{n+n^\prime}[\Gamma(n)\Gamma(n^\prime)]^{1/2}\,\nu!\nu^\prime!}
{[\Gamma(n+l+\frac{1}{2})\Gamma(n^\prime+l^\prime+\frac{1}{2})]^{1/2}}\times\\
&&\sum_{\sigma}\frac{\Gamma(t+\sigma)}{(\sigma-1)!(n-\sigma)!(n^\prime-\sigma)!
(\sigma+\nu-n)!(\sigma+\nu^\prime-n)!},\nonumber
\eeqn%
with the integers $t=\frac{1}{2}(l+l^\prime+\lambda+1)$,
$\nu=\frac{1}{2}(l^\prime-l+\lambda)$, and
$\nu^\prime=\frac{1}{2}(l-l^\prime+\lambda)$.

Apart from parity conservation ($l+l^\prime+\lambda \equiv$ even),
the angular part does not depend explicitly on orbital angular momenta:
 \beqn
 \langle ljm\vert Y_{\lambda\mu}\vert l^\prime j^\prime m^\prime\rangle
 &=&(-1)^{j-m}\left(\begin{array}{ccc}
 j & \lambda& j^\prime \\
 -m & \mu   & m^\prime
 \end{array}\right)
 \langle j\vert\vert Y_\lambda\vert\vert j^\prime\rangle \; , \nonumber\\
 \eeqn
where the irreducible matrix elements of the spherical harmonic are given by
the expression
 \beq
 \langle j\vert\vert Y_\lambda\vert\vert j^\prime\rangle
 =(-1)^{j-\frac{1}{2}}\sqrt{\frac{\hat{j}\hat{j^\prime}\hat{\lambda}}{4\pi}}
 \left(\begin{array}{ccc}
 j         & \lambda &j^\prime\\
 -\frac{1}{2} & 0       & \frac{1}{2}
 \end{array}
\right) \; .%
\eeq %
\end{appendix}

\end{document}